\documentclass[twocolumn]{aastex61}
\usepackage{graphicx}
\usepackage{amsmath, amssymb, amsfonts}
\usepackage{epstopdf}
\usepackage{xspace}
\usepackage{natbib}
\usepackage{subfigure}
\usepackage{color}

\newcommand{\hst}{\textit{HST}\xspace}
\newcommand{\lenstool}{{\tt Lenstool}\xspace}

\newcommand{\spt}{SPT0615\xspace}
\newcommand{\aic}{AICc\xspace}

\submitjournal{ApJ}

\shorttitle{SPT-CLJ0615$-$5746}
\shortauthors{Paterno-Mahler et al.}

\turnoffeditone

\begin{document}

\title{RELICS: A Strong Lens Model for SPT-CLJ0615$-$5746, a $z=0.972$ Cluster}

\correspondingauthor{Rachel Paterno-Mahler}
\email{rachelpm@umich.edu}

\author{Rachel Paterno-Mahler}
\affil{Department of Physics and Astronomy, University of California, Irvine, 4129 Frederick Reines Hall, Irvine, CA 92697, USA}
\affil{Department of Astronomy, University of Michigan, 1085 South University Drive, Ann Arbor, MI 48109, USA}

\author{Keren Sharon}
\affil{Department of Astronomy, University of Michigan, 1085 South University Drive, Ann Arbor, MI 48109, USA}

\author{Dan Coe}
\affil{Space Telescope Science Institute, 3700 San Martin Drive, Baltimore, MD 21218, USA}

\author{Guillaume Mahler}
\affil{Department of Astronomy, University of Michigan, 1085 South University Drive, Ann Arbor, MI 48109, USA}

\author{Catherine Cerny}
\affil{Department of Astronomy, University of Michigan, 1085 South University Drive, Ann Arbor, MI 48109, USA}
\affil{Astronomy Department and Institute for Astrophysical Research, Boston University, 725 Commonwealth Avenue, Boston, MA 02215, USA}

\author{Traci L. Johnson}
\affil{Department of Astronomy, University of Michigan, 1085 South University Drive, Ann Arbor, MI 48109, USA}

\author{Tim Schrabback}
\affil{Argelander-Institut f{\"u}r Astronomie, Universit{\"a}t Bonn, Auf dem H{\"u}gel 71, 53121, Bonn, Germany}

\author{Felipe Andrade-Santos}
\affil{Harvard-Smithsonian Center for Astrophysics, 60 Garden Street, Cambridge, MA 02138, USA}

\author{Roberto J. Avila}
\affil{Space Telescope Science Institute, 3700 San Martin Drive, Baltimore, MD 21218, USA}

\author{Maru\v{s}a Brada\v{c}}
\affil{Department of Physics, University of California, Davis, CA 95616, USA}

\author{Larry D. Bradley}
\affil{Space Telescope Science Institute, 3700 San Martin Drive, Baltimore, MD 21218, USA}


\author{Daniela Carrasco}
\affil{School of Physics, University of Melbourne, VIC 3010, Australia}

\author{Nicole G. Czakon}
\affil{Institute of Astronomy and Astrophysics, Academia Sinica, PO Box 23-141, Taipei 10617,Taiwan}

\author{William A. Dawson}
\affil{Lawrence Livermore National Laboratory, P.O. Box 808 L-210, Livermore, CA, 94551, USA}

\author{Brenda L. Frye}
\affil{Department of Astronomy, Steward Observatory, University of Arizona, 933 North Cherry Avenue, Rm N204, Tucson, AZ, 85721, USA}

\author{Austin T. Hoag}
\affil{Department of Physics, University of California, Davis, CA 95616, USA}

\author{Kuang-Han Huang}
\affil{Department of Physics, University of California, Davis, CA 95616, USA}

\author{Christine Jones}
\affil{Harvard-Smithsonian Center for Astrophysics, 60 Garden Street, Cambridge, MA 02138, USA}

\author{Daniel Lam}
\affil{Leiden Observatory, Leiden University, NL-2300 RA Leiden, The Netherlands}

\author{Rachael Livermore}
\affil{School of Physics, University of Melbourne, VIC 3010, Australia}
\affil{ARC Centre of Excellence for All Sky Astrophysics in 3 Dimensions (ASTRO 3D), VIC 2010, Australia}

\author{Lorenzo Lovisari}
\affil{Harvard-Smithsonian Center for Astrophysics, 60 Garden Street, Cambridge, MA 02138, USA}

\author{Ramesh Mainali}
\affil{Department of Astronomy, Steward Observatory, University of Arizona, 933 North Cherry Avenue, Rm N204, Tucson, AZ, 85721, USA}

\author{Pascal A. Oesch}
\affil{Department of Astronomy, University of Geneva, Chemin des Maillettes 51, 1290 Versoix, Switzerland}

\author{Sara Ogaz}
\affil{Space Telescope Science Institute, 3700 San Martin Drive, Baltimore, MD 21218, USA}

\author{Matthew Past}
\affil{Department of Astronomy, University of Michigan, 1085 South University Drive, Ann Arbor, MI 48109, USA}

\author{Avery Peterson}
\affil{Department of Astronomy, University of Michigan, 1085 South University Drive, Ann Arbor, MI 48109, USA}



\author{Russell E. Ryan}
\affil{Space Telescope Science Institute, 3700 San Martin Drive, Baltimore, MD 21218, USA}

\author{Brett Salmon}
\affil{Space Telescope Science Institute, 3700 San Martin Drive, Baltimore, MD 21218, USA}

\author{Irene Sendra-Server}
\affil{Infrared Processing and Analysis Center, California Institute of Technology, MS 100-22, Pasadena, CA 91125}

\author{Daniel P. Stark}
\affil{Department of Astronomy, Steward Observatory, University of Arizona, 933 North Cherry Avenue, Rm N204, Tucson, AZ, 85721, USA}



\author{Keiichi Umetsu}
\affil{Institute of Astronomy and Astrophysics, Academia Sinica, PO Box 23-141, Taipei 10617,Taiwan}

\author{Benedetta Vulcani}
\affil{School of Physics, University of Melbourne, VIC 3010, Australia}

\author{Adi Zitrin}
\affil{Physics Department, Ben-Gurion University of the Negev, P.O. Box 653, Beer-Sheva 84105, Israel}

\begin{abstract}

We present a lens model for the cluster SPT-CLJ0615$-$5746, which is the highest redshift ($z=0.972$) system in the Reionization of Lensing Clusters Survey (RELICS), making it the highest redshift cluster for which a full strong lens model is published.  We identify three systems of multiply-imaged lensed galaxies, two of which we spectroscopically confirm at $z=1.358$ and $z=4.013$, which we use as constraints for the model.  We find a foreground structure at $z\sim0.4$, which we include as a second cluster-sized halo in one of our models; however two different statistical tests find the best-fit model consists of one cluster-sized halo combined with three individually optimized galaxy-sized halos, as well as contributions from the cluster galaxies themselves.  We find the total projected mass density within $r=26.7''$ (the region where the strong lensing constraints exist) to be $M=2.51^{+0.15}_{-0.09}\times 10^{14}$~M$_{\sun}$.  If we extrapolate out to $r_{500}$, our projected mass density is consistent with the mass inferred from weak lensing and from the Sunyaev-Zel'dovich effect ($M\sim10^{15}$~M$_{\sun}$).  This cluster is lensing a previously reported $z\sim10$ galaxy, which, if spectroscopically confirmed, will be the highest-redshift strongly lensed galaxy known.   

\end{abstract}

\keywords{galaxies:clusters:individual (SPT-CLJ0615$-$5746)--gravitational lensing:strong}

\section{Introduction} \label{intro}


Gravitational lensing occurs when light from a background object is deflected around mass between the object and the observer.  The amount of deflection is related to the strength of the gravitational field; i.e., the mass distribution, as well as to the geometrical configuration of the lens, source, and observer.  The deflection is independent of the type of matter and its state, meaning that lensing is sensitive to both luminous and dark matter.  Thus, it is ideal for measuring the projected mass density of the cluster core to great precision out to the location of the strong lensing constraints.  Nevertheless, strong and weak lensing measurements of mass and lensing magnifications are prone to systematic uncertainties~\citep{johnson, meneghetti}.  Most notably, it is sensitive to structure along the line of sight (e.g., \citealp{daloisio, bayliss, jaro, mccully, chirivi}), as all matter along the line of sight contributes to the observed lensing signal.

While there are quite a few known strong lensing clusters at lower redshifts, there are only a handful at $z>0.8$, despite the many targeted searches for high redshift clusters~\citep{w14, bleem, cobra}.   For many of these high-redshift strong-lensing clusters, strongly lensed galaxies are observed in the form of stretched arcs; however no detailed lens models exist in the literature~\citep{huang, gonzalez}.  This is likely due to the difficulties in computing such models: they require a large investment of time on the \textit{Hubble} Space Telescope (\hst) to obtain enough constraints, as well as spectroscopic follow-up to obtain redshifts. 

Mass modeling of strong gravitational lenses at a large range of redshifts allows us to test predictions about the universe.  We can compare the observed distribution of lenses, lens mass, and the distribution of the brightness of lensed galaxies (among other properties) to simulations for varying cosmological parameters to test our theories.  Such studies have been done for small cluster samples~\citep{bartelmann, wambsganss, dalal, ho, li, sand, hennawi, horesh, bayliss2011, xu}.

Here we present a strong lens model for the cluster SPT-CLJ0615$-$5746 (also known as PLCKG266.6$-$27.3; hereafter \spt; RA: 06h15m56s, DEC: $-57^\circ 45' 50''$; \citealp{planck, williamson, bleem}).  This is the highest redshift cluster in the Reionization of Lensing Clusters Survey (RELICS) sample, with $z=0.972$~\citep{sz2}.  The study of lensing clusters in the $z\sim1-2$ regime is crucial to understanding the statistics described above, as some of the lensed galaxies behind high-redshift lensing clusters should not exist due to their brightness, based on current realistic assumptions~\citep{gonzalez}.  A statistical sample of high-redshift lensing clusters give us the ability to understand the true frequency of lensed galaxies behind high-redshift clusters. 

The goal of the RELICS project is to find a statistically significant sample of galaxies at high redshift to constrain the luminosity function at $z>6$~\citep{salmon} and probe the epoch of reionization at $z>9$~\citep{arc}.  RELICS uses gravitational lensing by galaxy clusters to search for these magnified high-redshift galaxies; secondary science goals include cluster physics (such as mass scaling relations) and discovering supernovae.  Archival HST imaging reveals that \spt is a strong lensing cluster.  The primary lensing evidence comes from a source galaxy nearly directly behind the cluster is strongly lensed into three images, which are the most notable strong lensing constraints in the field.  We use these, along with other newly discovered lensed galaxies and their spectroscopic redshifts, to determine a strong lensing mass model of \spt.

This paper is organized as follows: in \S\ref{data} we present the data from the various observatories used and in \S\ref{models} we present our modeling efforts.  In \S\ref{conc} we discuss the results of our modeling and compare our results to other high-redshift clusters that also have strong lens models.  Throughout this work we assume a flat cosmology with $H_0=70$~km~s$^{-1}$~Mpc$^{-1}$, $\Omega_{\Lambda}=0.7$, and $\Omega_{M}=0.3$.  At the redshift of \spt ($z=0.972$), this gives a scale of $1\arcsec=7.953$~kpc and a luminosity distance of $D_L=6379.3$~Mpc.  \edit1{We adopt the standard notation of $M_{\Delta}$ to denote the mass enclosed within a sphere of radius $r_{\Delta}$, within which the mean overdensity equals $\Delta$ times the critical density of the universe at the cluster redshift, $z=0.972$.}

\section{Data and Data Reduction} \label{data}
\subsection{HST Imaging} \label{hst}
\spt was observed with \hst as part of the Reionization of Lensing Clusters Survey (RELICS, GO-14096, PI: Coe) Treasury \hst program, which aimed to discover a statistically significant samples of galaxies at high redshift ($z>6$, \citealp{salmon}).  The cluster selection process is described in detail in ~\citet{cerny} and Coe et al. (in prep), and strong lensing analyses for other RELICS clusters were published in \citet{cerny}, \citet{acebron}, and \citet{cibirka}.  \spt was observed for two orbits with the Wide Field Camera 3 (WFC3) in F105W, F125W, F140W, F160W and for one orbit with the Advanced Camera for Survey (ACS) in F435W.  All clusters in the program were imaged over two epochs to allow for variability searches. Additional archival ACS imaging in F606W and F814W were available from GO-12757 (PI: High) and GO-12477 (PI: Mazzotta).  GO-12477 obtained one pointing of F814W imaging and a $2\times2$ mosaic in F606W.  GO-12757 obtained a $2\times2$ mosaic in F814W, including overlapping area for deeper imaging in the strong lensing region. \edit1{Because the length of the exposure time varies in F814W, the depth of the field varies from $m_{814,AB}=27.99$ at the very center of the field to $m_{814,AB}=27.19$ at the shallowest part of the field.  These are the 5-sigma limiting magnitudes in a $r=0\farcs2$ circular aperture.} The center of the field will have a deeper limiting magnitude.  The wavelength coverage spans $0.4-1.7$~\micron.  Table~\ref{obslog} summarizes the observations. 

Calibrated images from all available programs, including archival programs, were obtained from the Mikulski Archive for Space Telescopes (MAST)\footnote{https://archive.stsci.edu}.  Individual frames were then visually inspected to ensure that the quality is acceptable for science.  Satellite trails and other image artifacts were manually masked out.  Additionally, the WFC3/IR images have persistence which was masked out using products supplied by the WFC3 team.  A custom pixel mask provided by G. Brammer (personal communication) removes hot pixels not in the pipeline mask.  The ACS images were corrected for charge transfer inefficiency losses using the method described in ~\citet{cte}.  Sub-exposures in each filter were combined to form a deep image using the AstroDrizzle package~\citep{drizzlepac} using \texttt{PIXFRAC}$=0.8$.  The images in different filters were aligned to the same reference frame, and the astrometry was matched to the Wide-field Infrared Survey Explorer (WISE) point source catalog~\citep{wise}.  The final, reduced images are made available to the public as high level science products through MAST\footnote{https://archive.stsci.edu/prepds/relics}.  The public release includes photometric catalogs of all the fields, including photometric redshift estimates using the Bayesian Photometric Redshifts method (BPZ; \citealp{bpz}).

\begin{deluxetable*}{clcl}
\tablecaption{Observation Information \label{obslog}}
\tablecolumns{4}
\tablewidth{0pt}
\tablehead{
\colhead{Instrument} &
\colhead{Exp. Time (s)} &
\colhead{UT Date} &
\colhead{Program}
}
\startdata
ACS/WFC F435W&2249&2017-02-08&GO14096\tablenotemark{a}\\
ACS/WFC F606W&1920&2012-01-20&GO12477\tablenotemark{b}\\
ACS/WFC F606W&1920&2012-01-20&GO12477\tablenotemark{b}\\
ACS/WFC F606W&1920&2012-01-21&GO12477\tablenotemark{b}\\
ACS/WFC F606W&1920&2012-01-21&GO12477\tablenotemark{b}\\
ACS/WFC F814W&2476&2012-01-19&GO12757\tablenotemark{b}\\
ACS/WFC F814W&2476&2012-01-19&GO12757\tablenotemark{b}\\
ACS/WFC F814W&1916&2012-01-21&GO12477\\
ACS/WFC F814W&2476&2012-01-22&GO12757\tablenotemark{b}\\
ACS/WFC F814W&2476&2012-01-25&GO12757\tablenotemark{b}\\
WFC3/IR F105W&755.9&2017-02-08&GO14096\tablenotemark{a}\\
WFC3/IR F105W&755.9&2017-03-23&GO14096\tablenotemark{a}\\
WFC3/IR F125W&380.9&2017-02-08&GO14096\tablenotemark{a}\\
WFC3/IR F125W&380.9&2017-03-23&GO14096\tablenotemark{a}\\
WFC3/IR F140W&380.9&2017-02-08&GO14096\tablenotemark{a}\\
WFC3/IR F140W&380.9&2017-03-23&GO14096\tablenotemark{a}\\
WFC3/IR F160W&1055.9&2017-02-08&GO14096\tablenotemark{a}\\
WFC3/RI F160W&1055.9&2017-03-23&GO14096\tablenotemark{a}\\
\enddata
\tablenotetext{a}{RELICS program}
\tablenotetext{b}{These images are different pointings of a $2\times2$ mosaic.}
\end{deluxetable*}

\subsection{Ground-Based Spectroscopy} \label{spec}

Ground-based spectroscopic observations were obtained using the upgraded Low Dispersion Survey Spectrograph (LDSS3-C) on the Magellan Clay telescope using University of Arizona (PI: Stark) allocation.  \spt was observed on 2017 March 30 for a total exposure time of one hour.  Average seeing was $0\farcs6-0\farcs7$ throughout the night.  Slits were placed on candidate lensed galaxies.  The VPH-ALL grism was used, which has coverage between $4250~\textrm{\AA} < \lambda < 10000~\textrm{\AA}$.  A 1\arcsec\ slit was used on all objects, with spectral resolution R~450-1100 across the wavelength range.  The detector is  $6\farcm4$ in spatial extent.  A full description of the RELICS Magellan/LDSS3 followup results will be presented in a future paper (Mainali et al. in prep).

\section{Lens Model} \label{models}

The model is computed using \lenstool~\citep{lenstool}, which is a parametric model that uses Monte Carlo Markov Chain (MCMC) analysis to sample the parameter space.  Each dark matter halo is modeled as a pseudo-isothermal ellipsoidal mass distribution (PIEMD; \citealp{piemd}) with seven parameters: position (RA, DEC), mass (or velocity dispersion, $\sigma$), ellipticity ($\epsilon$), position angle ($\theta$), core radius ($r_{core}$), and truncation radius ($r_{cut}$). Dark matter halos are assigned to both the cluster as a whole and to individual cluster galaxies.  Cluster galaxies are selected via the cluster red sequence~\citep{rs}.  The position and shape parameters of cluster galaxies are fixed to their observed properties as measured from the galaxy light using \textit{Source Extractor}~\citep{sextractor}, and their mass-to-light ratios are assigned using scaling relations~\citep{piemd}.   The parameters for the cluster halos are allowed to vary, with the exception of the truncation radius that lies far beyond the strong lensing projected radius and thus cannot be constrained by the lensing evidence.  The truncation radius was fixed to 1500~kpc.

\begin{figure*}
\begin{center}
\plotone{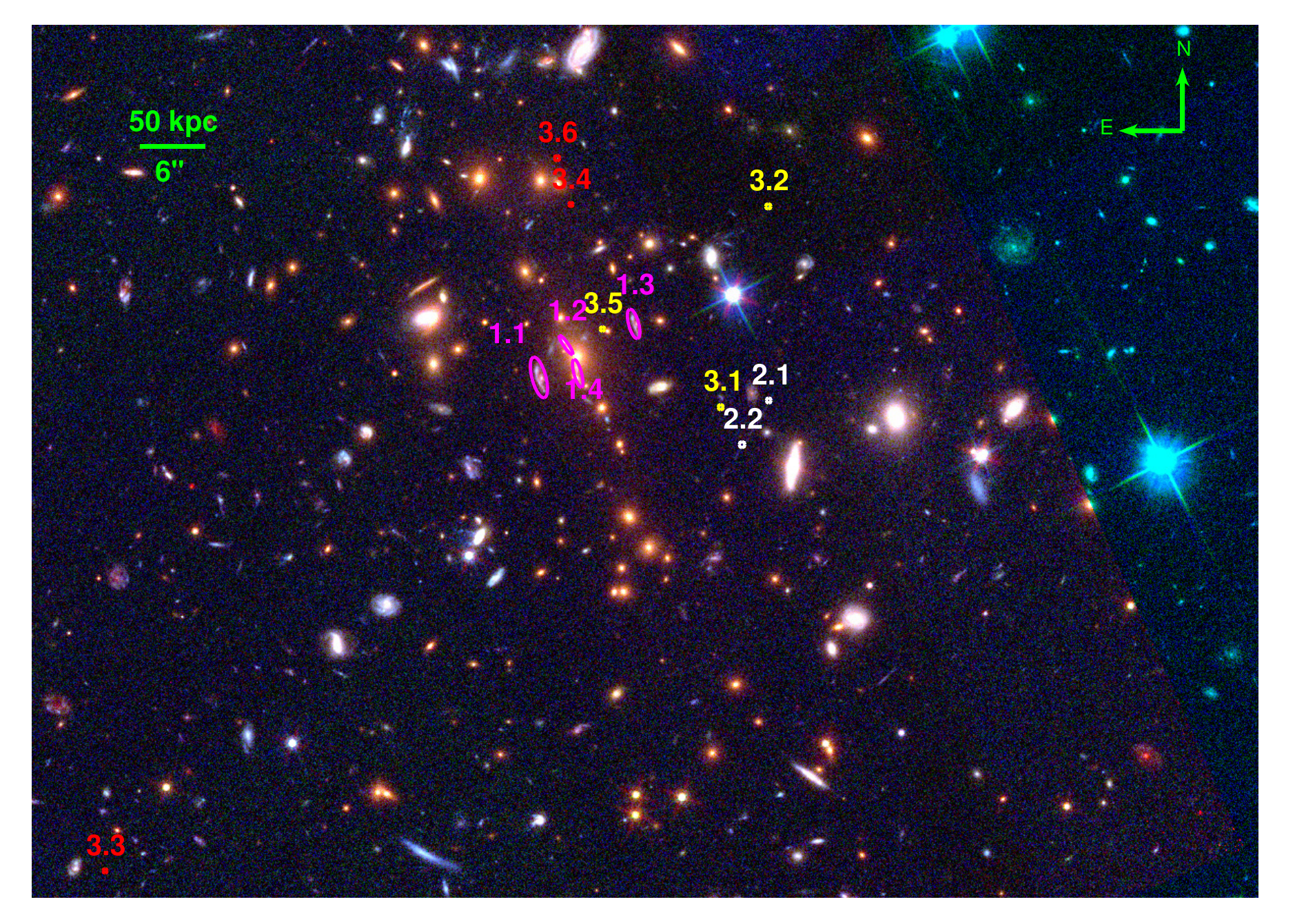}
\caption{Multiply imaged systems used in the lens model on a composite WFC3/IR F160, ACS F814, and ACS F606 HST image of \spt.  System 1 has a spectroscopically determined redshift of $z=1.358$ and is shown in purple.  For clarity, the individual sub-systems are not labeled.  System 2 is shown in white.  System 3 has a spectroscopically determined redshift of $z=4.013$.  Images used in models 1 and 3 are shown in yellow.  These are the most secure detections.  Two of the three (3.1 and 3.2) are spectroscopically confirmed.  Models 2 and 4 include all constraints in system 3.}
\label{arcs}
\end{center}
\end{figure*}

\begin{deluxetable*}{lcccccccccccccc}
\rotate
\setlength{\tabcolsep}{0.04in}
\tablecolumns{14}
\tablecaption{Properties of Lensed Galaxies\label{table:arcs}}
\tablewidth{0pt}
\tablehead{
\colhead{ID} &
\colhead{RA} &
\colhead{DEC} &
\colhead{RELICS ID\tablenotemark{a}} &
\colhead{Photo-$z$ [$z_{min}$,$z_{max}$]} &
\colhead{Spec-$z$} &
\colhead{M1-$z$} &
\colhead{M1 rms} &
\colhead{M2-$z$} &
\colhead{M2 rms} &
\colhead{M3-$z$} &
\colhead{M3 rms} &
\colhead{M4-$z$} &
\colhead{M4 rms}
}
\startdata
1.1 & 06 15 52.22 & $-$57 46 49.9 & 613 & 1.23 [1.16,1.31] & 1.358 & \nodata & 0.13 & \nodata & 0.70 & \nodata & 0.13 & \nodata & 0.13\\
1.2 & 06 15 51.87 & $-$57 46 46.9  &\nodata& \nodata &  &  &  &  &  &  &  &  & \\
1.3 & 06 15 51.05& $-$57 46 44.7 & 631 & 1.16 [1.09,1.25] &  &  &  &  &  &  &  &  & \\
\hline 
10.1 & 06 15 52.15 & $-$57 46 50.6 & 614 & 1.18 [1.13,1.26] & 1.358 & \nodata & 1.05 & \nodata & 0.79 & \nodata & 0.80 & \nodata & 0.83\\
10.2 & 06 15 51.83 & $-$57 46 47.7 & \nodata &  &  &  &  &  &  &  &  & \\
10.3 & 06 15 50.99 & $-$57 46 45.3 & 632 & 1.29 [1.21,1.38] &  &  &  &  &  &  &  &  & \\
10.4 & 06 15 51.73 & $-$57 46 49.6 & \nodata &  &  &  &  &  &  &  &  & \\
\hline
11.1 & 06 15 52.17 & $-$57 46 51.0 & 614 & 1.18 [1.13,1.26] & 1.358 & \nodata & 1.23 & \nodata & 1.20 & \nodata & 1.17 & \nodata & 1.06\\
11.2 & 06 15 51.79 & $-$57 46 47.8 & \nodata & \nodata &  &  &  &  &  &  &  &  & \\
11.3 & 06 15 51.00 & $-$57 46 45.7 & \nodata & \nodata &  &  &  &  &  &  &  &  & \\
11.4 & 06 15 51.72 & $-$57 46 49.8 & \nodata & \nodata &  &  &  &  &  &  &  &  & \\
\hline
12.1 & 06 15 52.11 & $-$57 46 51.8 & 614 & 1.18 [1.13,1.26] & 1.358 & \nodata & 0.13 & \nodata & 0.38 & \nodata & 0.22 & \nodata & 0.09\\
12.3 & 06 15 50.99 & $-$57 46 46.5 & 632 & 1.29 [1.21,1.38] &  &  &  &  &  &  &  &  & \\
12.4 & 06 15 51.66 & $-$57 46 51.1 & \nodata & 1.04 & \nodata &  &  &  &  &  &  &  &  & \\
\hline
2.1 & 06 15 49.37 & $-$57 46 52.8 & 729 & 0.79 [0.20,3.80] & \nodata & $2.43^{+0.07}_{-0.12}$ & 0.43 & $2.10^{+0.03}_{-0.07}$ & 0.15 & $2.48^{+0.01}_{-0.27}$ & 0.19 & $2.30^{+0.07}_{-0.08}$ & 0.06\\
2.2 & 06 15 49.70 & $-$57 46 57.1 & \nodata & 2.7\tablenotemark{b} &  &  &  &  &  &  &  &  & \\
\hline
3.1 & 06 15 49.96 & $-$57 46 53.5 & 725 & 4.16 [4.02,4.25] & 4.013 & \nodata & 0.22 & \nodata & 1.99 & \nodata & 0.26 & \nodata & 0.35\\
3.2 & 06 15 49.38 & $-$57 46 33.9 & 494 & 4.26 [4.14,4.35] &  &  &  &  &  &  &  &  & \\
3.3 & 06 15 57.45 & $-$57 47 38.6 & 1196 & 4.16 [0.44,4.42] &  &  &  &  &  &  &  &  & \\
3.4 & 06 15 51.78 & $-$57 46 33.7 & 493 & 4.22 [4.04,4.36] &  &  &  &  &  &  &  &  & \\
3.5 & 06 15 51.39 & $-$57 46 45.9 & \nodata & \nodata &  &  &  &  &  &  &  &  & \\
3.6 & 06 15 51.95 & $-$57 46 29.2 & 440 & 0.47 [0.15,4.12] &  &  &  &  &  &  &  &  & \\
\enddata
\tablenotetext{a}{RELICS ID is based on the IR detection.}
\tablenotetext{b}{No ID was found in the IR images; this redshift is an upper limit based on the detection of many segments in the combined ACS/IR image.}
\tablecomments{RA and DEC are J2000.  Not all subsystems were detected by SExtractor and thus not all subsystems have photometric redshifts.  Photometric redshift ranges represent the 95\% confidence interval.  Spectroscopic redshifts were held fixed during modeling.  The rms is measured in the image plane for each system of multiple images and is measured in arcseconds.  Models 1 and 3 do not include galaxies 3.3, 3.4, and 3.6.  M1, M2, M3, and M4 refer to models 1-4 (see \S\ref{models}).}
\end{deluxetable*}

\begin{figure*}
\begin{center}
\plotone{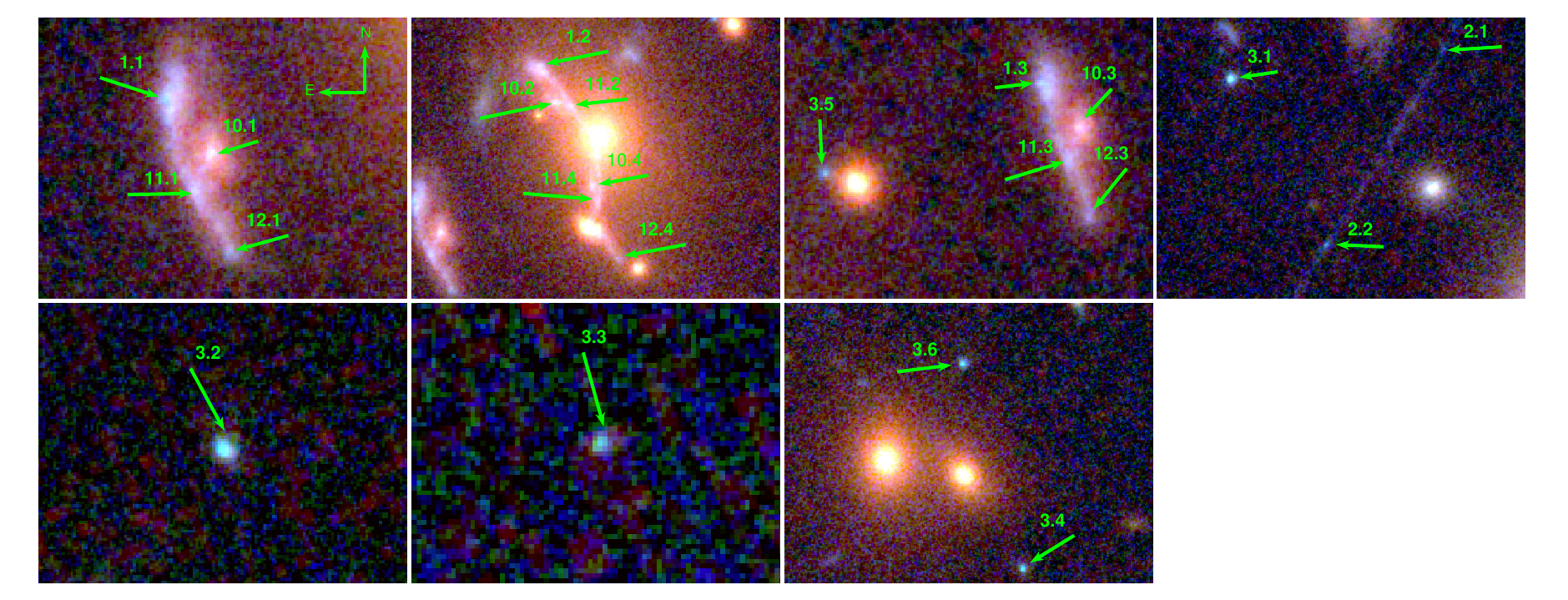}
\caption{Thumbnails of the individual systems described in the text.}
\label{thumbs}
\end{center}
\end{figure*}

For \spt, we identify three sets of multiply-imaged systems, shown in Figure~\ref{arcs}. \edit1{We note that there may be other lensing features in the field; however we did not use them as constraints because they were not confirmed as multiple images.  Additionally, there is a noticeable arced galaxy approximately $1''$ north of our high-redshift ($z\sim10$) candidate. Its photometric redshift is $z_{phot}\sim3$.  At this redshift it is not expected to be multiply imaged, and indeed our model does not predict any multiple images for this galaxy.  As such, we do not use it as an additional constraint.}  We show thumbnails of each image in Figure~\ref{thumbs}.  Their properties are described in Table~\ref{table:arcs}.  The constraints are identified by eye based on their morphology, structure, and color, and confirmed with the lens models. Using multi-object slit spectroscopy of this field using LDSS3 on the Magellan Clay telescope, we measure spectroscopic redshifts for two of the sources (for more information on the spectral observations, see Mainali et al. (in prep)).

\begin{figure*}
\begin{center}
\includegraphics[scale=0.55]{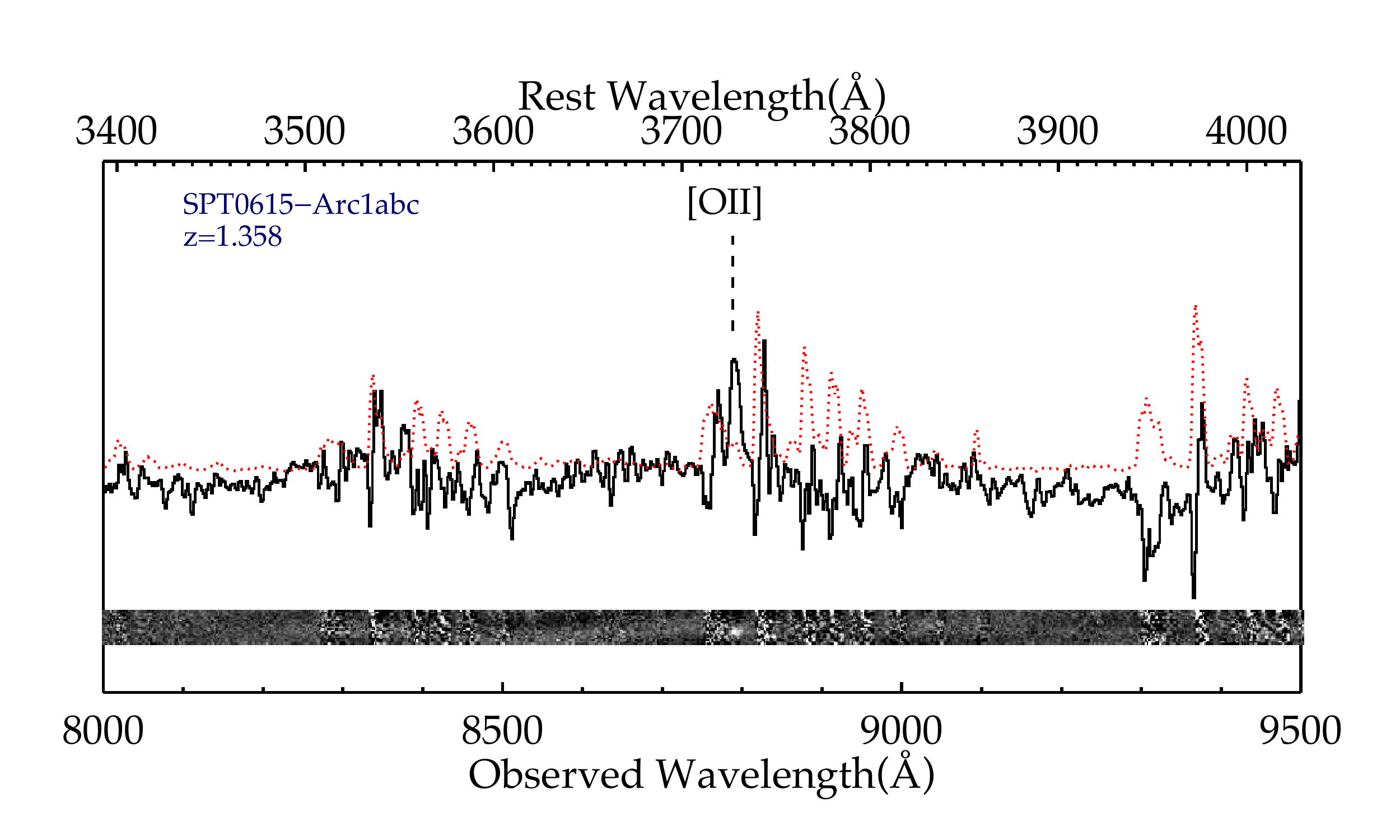}
\includegraphics[scale=0.6]{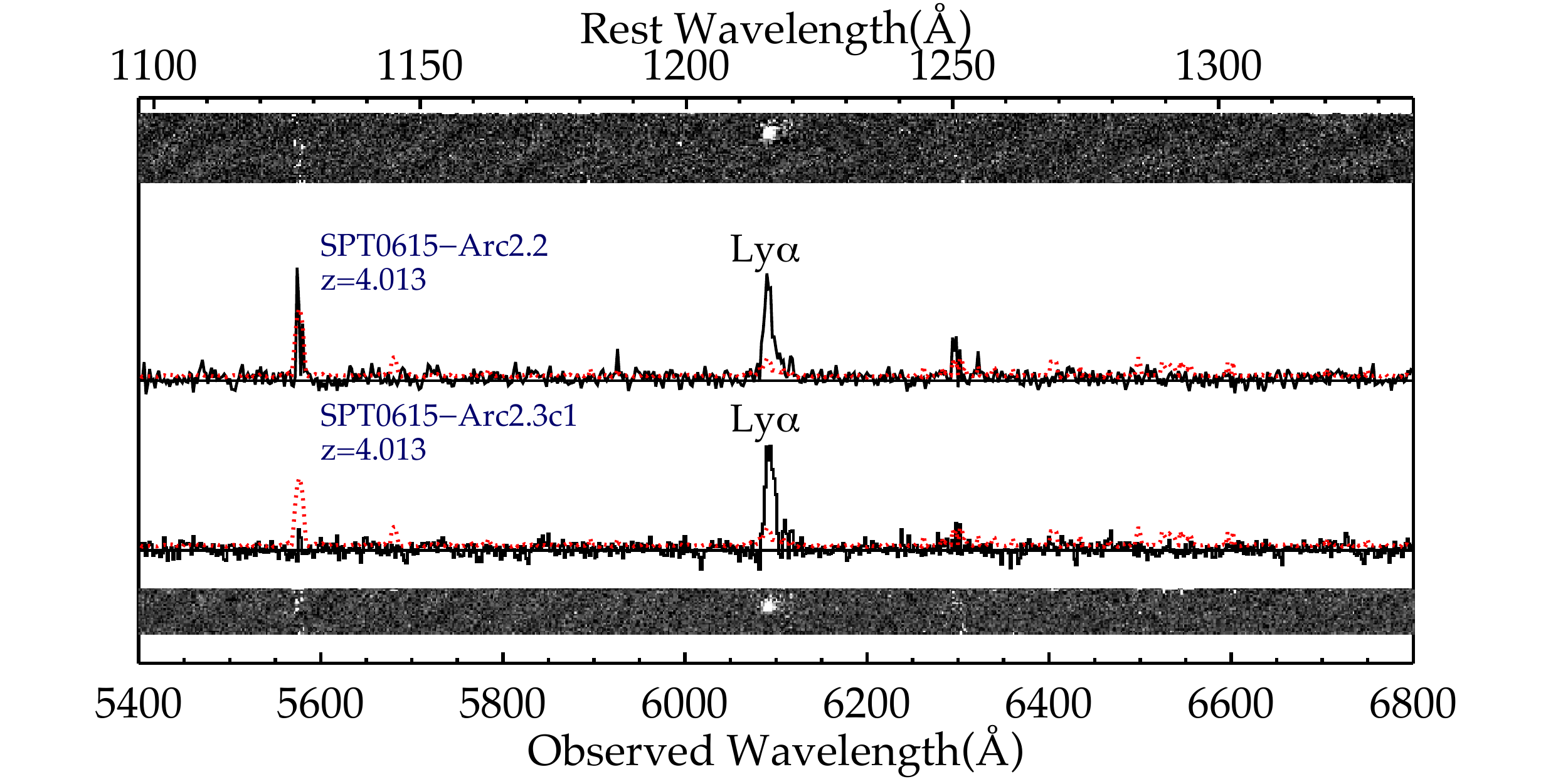}
\caption{Spectra used to determine the redshifts of system 1 (top) and system 3 (bottom). Each panel shows both the 2D and 1D spectrum, as well as the lines used to determine the redshift.  The solid black line is the spectrum of the object, while the dashed red line is the $1\sigma$ noise level (error spectrum).}
\label{spectra}
\end{center}
\end{figure*}

System 1 has a redshift of $z_{spec}=1.358$, determined by [OII] emission in image 1.1 (Figure~\ref{spectra}, top panel).  The galaxy has a distinctive shape, with four obvious knots. We use these knots as individual constraints.  All the images in this system are secure, as are each of the knots.  

System 2 consists of one long fold arc with mirror symmetry, with two secure detections.  Image 2.1 has a BPZ photometric redshift $z_{phot}=0.79$, with range $[0.20,3.80]$.  A single segment for image 2.2 could not be identified; however the different segments that comprise it has a maximum redshift of 2.7.  While the photometric redshifts of the two images in system 2 are disparate, the 95\% confidence interval on each is consistent and broad.  

System 3 is a compact galaxy at $z_{spec}=4.013$, determined with Ly-$\alpha$ emission (Figure~\ref{spectra}, bottom panel).  It is brightest in F814W, with a blue near-IR slope.  Slits were placed on both image 3.1 and image 3.2.  A redshift was measured from each slit placement.  Those, along with image 3.5, are secure identifications.  System 3 also has three other arc candidates that are less secure.  We explore the effect of adding those images to the model in more detail below.  We leave spectroscopically determined redshifts fixed during the modeling process.

In addition to the constraints discussed above, there is a candidate $z\sim10$ lensed galaxy in the field~\citep{arc}.  This candidate was not used as a constraint due to a lack of counter-images.  See \S\ref{hizarc} for more details on this galaxy.

\begin{figure*}
\begin{center}
\plottwo{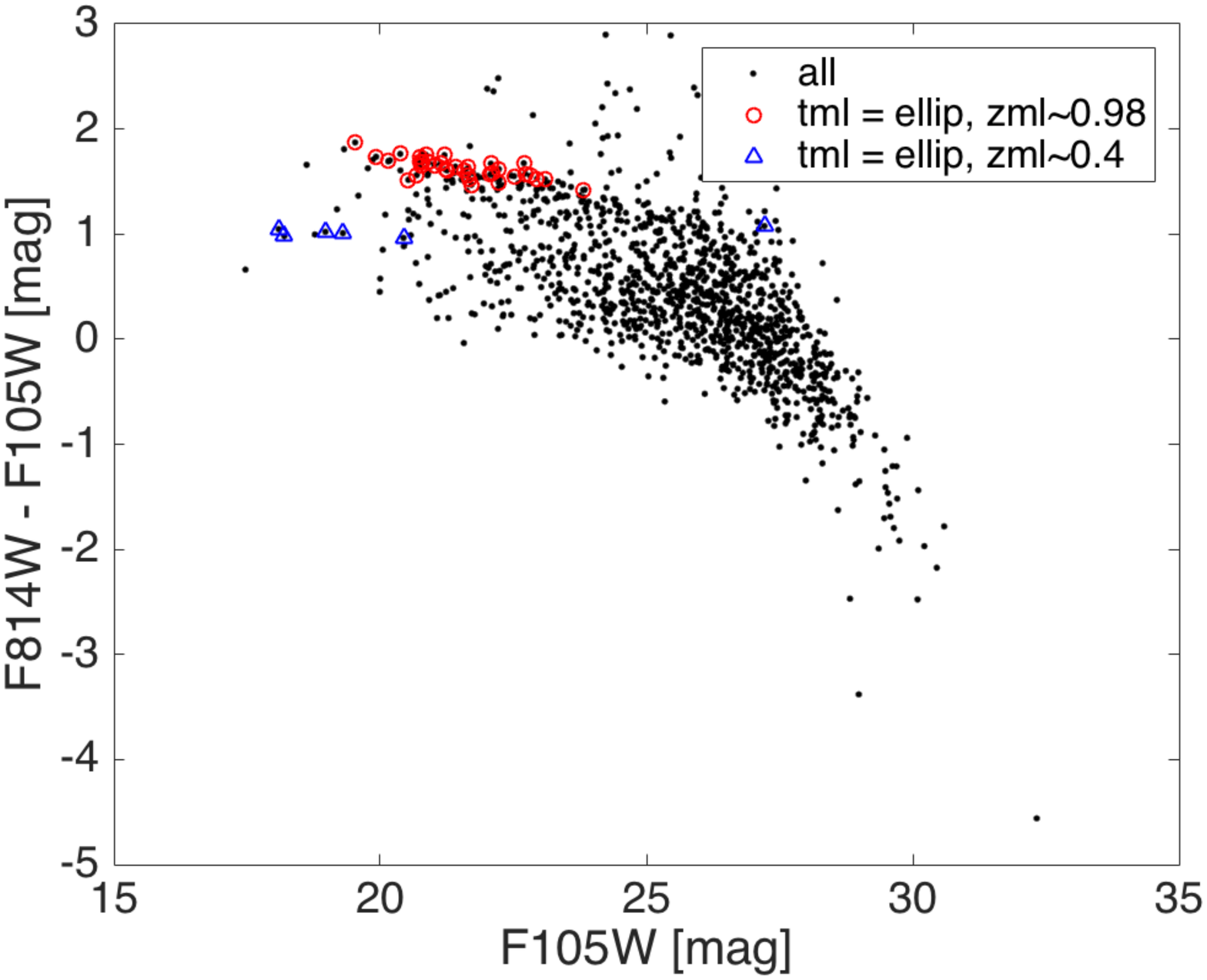}{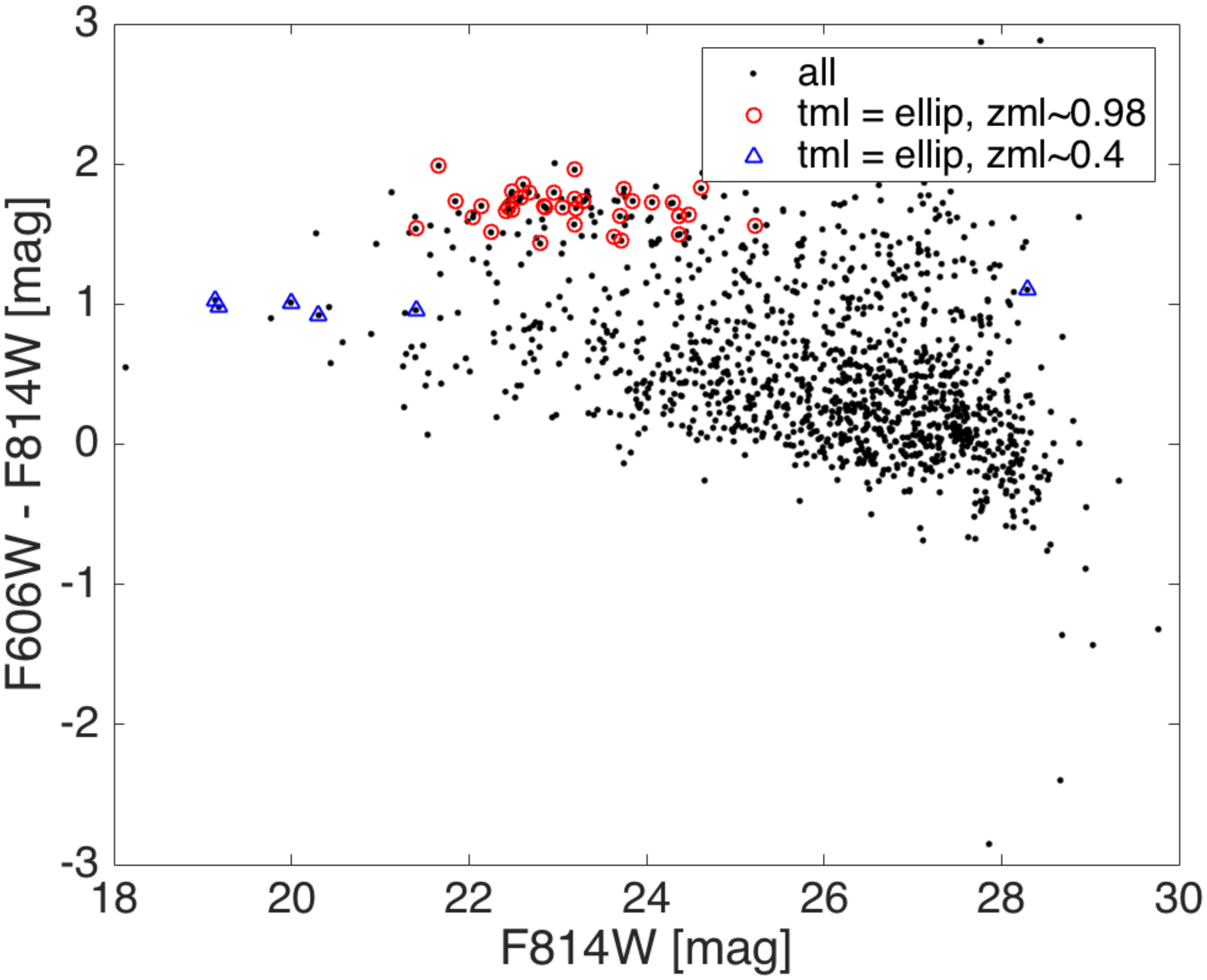}
\caption{Color-Magnitude diagrams showing the red sequence for both \spt at $z=0.972$ (red circles) and the foreground structure (blue triangles), estimated to be at $z\approx0.4$.  Both CMDs were created using photometric redshifts.  Black dots show all galaxies in the field, selected by their stellarity parameter.  Cluster galaxies and foreground structure galaxies were selected via their maximum likelihood most likely redshift (zml) and their maximum likelihood most likely spectral type (tml), \edit1{determined from the photometry in each filter and the BPZ templates.}  The left panel shows F814W$-$F105W plotted against F105W, which samples the galaxies of \spt better, while the right panel shows F606W$-$F814W plotted against F814W, which samples the galaxies of the foreground structure better.}
\label{cmd}
\end{center}
\end{figure*}

Figure~\ref{arcs} shows that there appears to be a foreground structure, with galaxies appearing bluer in color when compared with the color-selected galaxies of \spt.  In Figure~\ref{cmd}, we show the color-magnitude diagrams (CMDs) highlighting these two structures.  The main cluster forms an obvious red sequence, and there appears to be a second putative red sequence for a foreground structure at $z\approx0.4$, determined from the photometric redshifts of the members on the putative red sequence. 

Creating the model is an iterative process.  We start with one cluster-sized halo and an initial set of constraints, and add more halos and constraints until the model rms no longer improves.  While photometric redshifts exist for all of the lensed systems, we leave the redshifts of systems without a spectroscopically determined redshift free to vary during the modeling process so that it will not be affected by catastrophic outliers.  In \spt, the only system without a spectroscopically determined redshift is system 2.

Below we describe the four models that we consider, which take into account the various scenarios that can be applied to \spt.  As mentioned above, System 3 has three secure detections, along with three other multiple image candidates that were predicted by one of the models.  We create two different models, one with only the secure detections of system 3 and one with all of the detections of system 3, in order to compare them.  We also note that there is foreground structure, which is described above.  Because of this, we explore additional models that include the presence of a second cluster-sized halo at the redshift of \spt.  To determine the goodness-of-fit of each model, we employ two different statistical tests.  First, we compute the Bayesian Information Criterion (BIC, \citealp{bic}):
\begin{equation}
BIC = -2\ln{L} + k\ln{n}, 
\label{eq:bic}
\end{equation}
where $L$ is the maximum likelihood, $k$ is the number of free parameters, and $n$ is the number of constraints.  

The second test we consider is the corrected Aikake Information Criterion (\aic, \citealp{aicc, deriv}), which helps address the potential for overfitting:
\begin{equation}
AICc = 2k - 2\ln{L} + \frac{2k(k+1)}{n - k -1}.
\label{eq:aic}
\end{equation}
All terms are the same as in the BIC.

Both of these tests are used to evaluate the quality of the available models, and to assess the trade-off between the goodness-of-fit of the model and the complexity of the model.  The model with the lowest BIC is preferred.  To determine which model is the best using the AICc, the AICc values of each model are compared to the model with the lowest AICc value using the relative likelihood, $\exp{\left[(AICc_{min} - AICc_{i})/2\right]}$.  This is the likelihood that the $ith$ model minimizes information loss when compared to the model with the lowest AICc.

The results of the statistical tests for each model are displayed in Table~\ref{stats}.  The rms of each multiple image system in each model is displayed in Table~\ref{table:arcs}.

\subsection{Model 1:  One Lens Plane} \label{model1}

\begin{figure*}
\begin{center}
\plotone{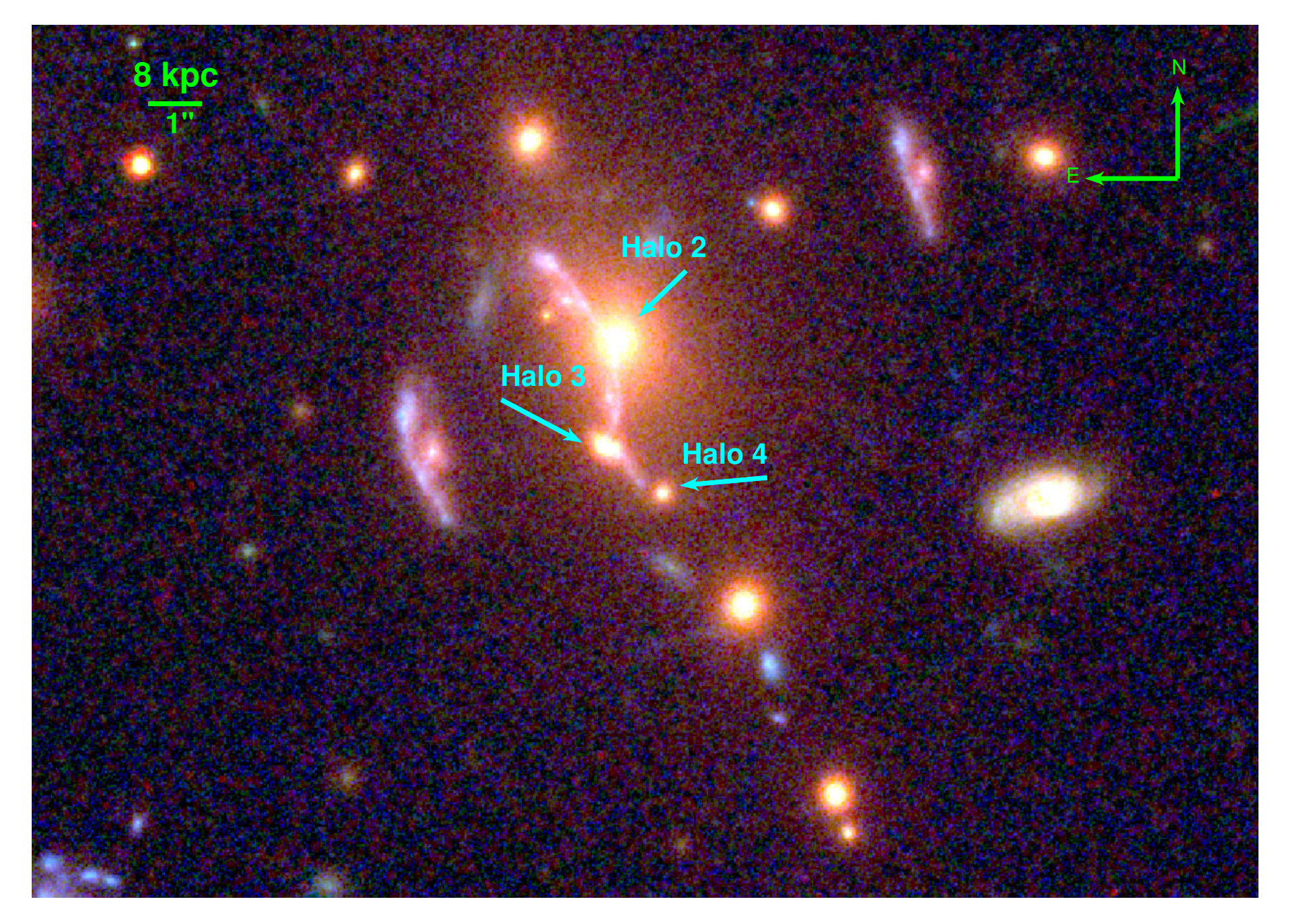}
\caption{Zoom in on the cluster center.  The individual galaxy halos that were allowed to vary are labeled in cyan.}
\label{zoom}
\end{center}
\end{figure*}

We first consider a model that includes all the images from systems 1 and 2, and three images from system 3.  This model has one cluster-sized halo and contributions from cluster-member galaxies as described above.  We fix the cut radius of this halo to 1500~kpc but allow all other parameters to vary.  Because of the proximity of the images in system 1 to the central cluster galaxies, we allow the velocity dispersion of three of the central cluster galaxies to vary (shown in Figure~\ref{zoom})  but fix all other parameters to those determined by scaling relations.  This is the model with the minimum BIC, -48.00, indicating that it is the best model by the standards of that criterion (see Table~\ref{stats}).  Compared to the other models, $\Delta BIC > 10$, meaning that the evidence in favor of this model is very strong.  It is also the best model using the \aic; none of the others are likely when compared to Model 1.  The critical curves for this model are shown \edit1{in Figure~\ref{fig:critic}}.  The model parameter results are shown in Table~\ref{params}.

\begin{deluxetable}{lllrrrr}
\tablecaption{Statistical Results \label{stats}}
\tablecolumns{7}
\tablewidth{0pt}
\tablehead{
\colhead{Model} &
\colhead{$n$} &
\colhead{$k$} &
\colhead{$\ln L$} &
\colhead{BIC} &
\colhead{\aic} &
\colhead{$\chi^2/d.o.f.$}
}
\startdata
1 & 32 & 11 & 43.06 & $-$48.00 & $-$50.92&11.73/15\\
2 & 38 & 11 & 7.19 & 25.63 & 17.77&96.17/21\\
3 & 32 & 17 & 46.35 & $-$33.78 & $-$14.99&8.61/9\\
4 & 38 & 17 & 49.37 & $-$36.90 & $-$34.14&6.59/15\\
\enddata
\end{deluxetable}  

\subsection{Model 2:  One Lens Plane, All of System 3} \label{model2}

Model 1 predicts three additional arc candidates in system 3.  Candidate 3.3 is predicted to be $\sim1$ magnitude fainter than arcs 3.1 and 3.2, but $\sim1.6$ magnitudes brighter than 3.5.  Candidate 3.4 is predicted to be 1.75 magnitudes brighter than arc 3.5.  Arc 3.1 has $m_{F814W}=25.52$ and Arc 3.2 has $m_{F814W}=25.49$.  Arc 3.5 could not be deblended from the neighboring source and thus we were unable to measure its magnitude.  Candidate 3.3 has $m_{F814W}=27.50$.  There are no predictions for the brightness of candidate 3.4 relative to arcs 3.1 and 3.2; however we measure its magnitude to be $m_{F814W}=26.40$.  Candidate 3.6 is predicted to be $0.1$ magnitudes fainter than arc 3.1 and $0.6$ magnitudes fainter than arc 3.2.  It is predicted to be 2.2 magnitudes brighter than arc 3.5.  We measure candidate 3.6 to be $m_{F814W}=25.70$.   We searched the regions of these predictions and found objects that were similar in color and morphology to the images with secure detections.  Model 2 includes all six of these images, but is otherwise the same as Model 1.  Table~\ref{stats} shows the results of the statistical tests.  Using both the BIC and \aic, this model is considered the worst or those tested.  The $\chi^2$ value for this model is also $\sim10\times$ higher than the $\chi^2$ value for any of the other models, and as such we do not consider it further, even taking into account the increased complexity of the model as compared to model 1.  As shown below, these constraints only make sense with a second halo to account for the foreground structure.  

\subsection{Model 3:  Foreground Structure} \label{model3}

In this model we attempt to account for the line-of-sight structure by adding a second cluster-sized halo to the single effective lens plane.  This line-of-sight structure is not associated with \spt,  so this is not a full multiplane analysis but rather an approximation.  Distance is degenerate with normalization, and with so few constraints it is difficult to disentangle the two.  This approximation ignores the higher order effects discussed in \citet{multiplane}, but does approximate the amplitude and direction of the shear that a second cluster-sized halo induces.  We fix the cut radius of this halo at 1800~kpc and allow all other parameters to vary.  The model puts this new halo directly to the south of the first cluster-sized halo.  The $\chi^2$ value for this model is comparable to that of \edit1{Models 1 and 4}.  It is the third most likely model of the four described here.

\subsection{Model 4:  Foreground Structure, All of System 3} \label{model4}

This model is the same as model 2 but adds an additional cluster-sized halo to account for the foreground structure.  As with model 3, we fix the cut radius of this second cluster-sized halo at 1800~kpc and allow all other parameters to vary.  If these three additional images are indeed part of system 3, as is indicated by their color and morphology, the separation is larger than expected in a typical lensing configuration, which could be caused by the presence of the foreground structure.  This is the second most probable model; however, using the relative likelihood estimator described above, it is only 0.02\% as likely as model 1 to be the best model.  The parameters of this model are shown in Table~\ref{params}.

\begin{figure*}
\begin{center}
\plotone{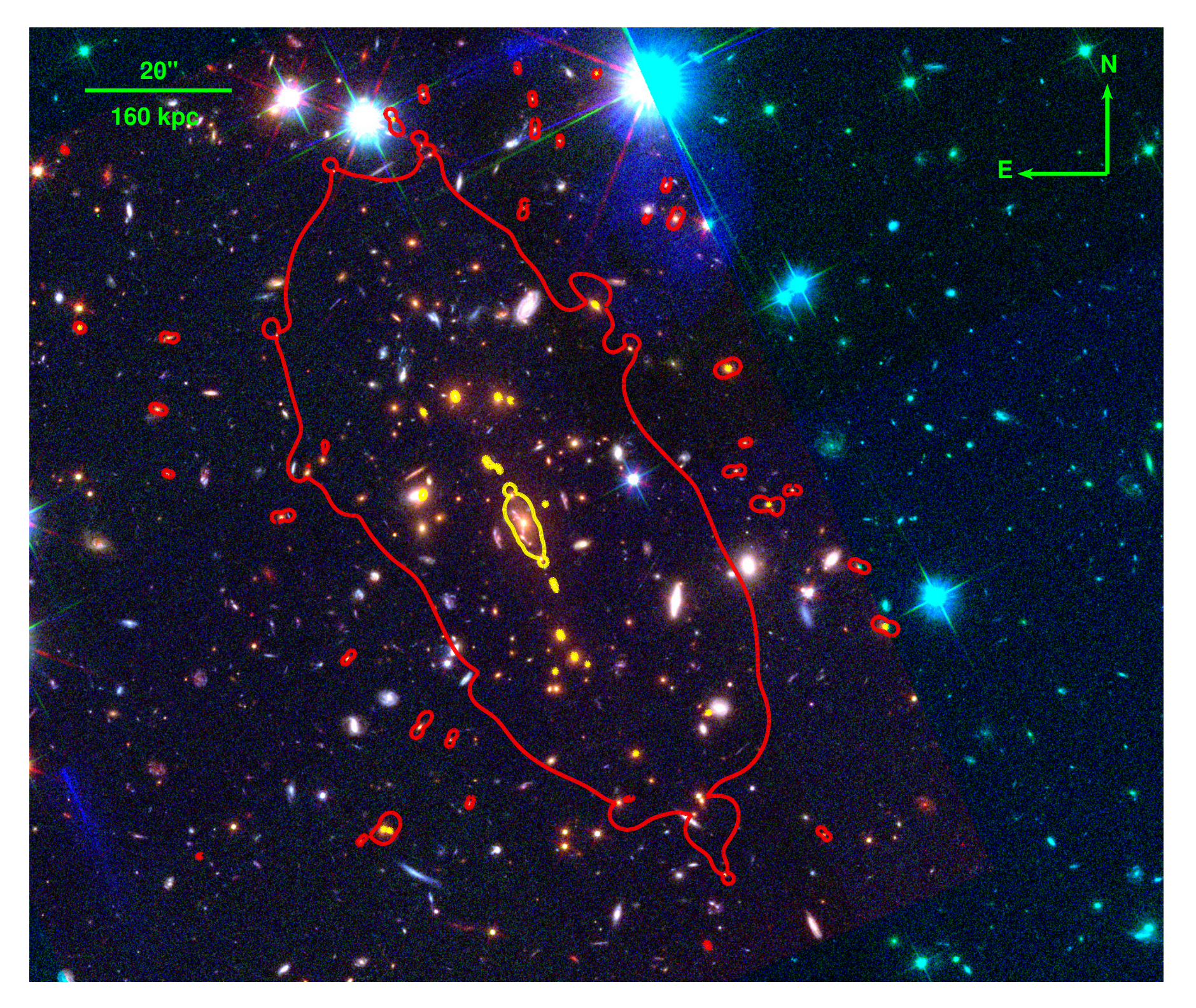}
\caption{Critical curves for Model 1 overlaid on a composite WFC3/IR F160, ACS F814, and ACS F606 HST image of \spt.  Critical curves for $z=1.3$ are in yellow, and the critical curves for $z=9.93$ (the redshift of the galaxy discussed in \S\ref{hizarc}) are in red.}
\label{fig:critic}
\end{center}
\end{figure*}

\begin{deluxetable*}{lrrrrrrr}
\tablecaption{Model Parameters \label{params}}
\tablecolumns{8}
\tablewidth{0pt}
\tablehead{
\colhead{Object} &
\colhead{$\Delta$ RA } &
\colhead{$\Delta$ DEC } &
\colhead{$\epsilon$} &
\colhead{$\theta$} &
\colhead{$r_{core}$} &
\colhead{$r_{cut}$} &
\colhead{$\sigma$}\\
\colhead{} &
\colhead{(kpc)} &
\colhead{(kpc)} &
\colhead{} &
\colhead{($^{\circ}$)} &
\colhead{($''$)} &
\colhead{($''$)} &
\colhead{(km~s$^{-1}$)}
}
\startdata
\cutinhead{Model 1}
Halo 1 &$0.40^{+0.41}_{-0.96} $& $3.63^{+1.16}_{-0.81}$ & $0.55^{+0.01}_{-0.05}$ & $124.2^{+1.4}_{-1.8}$ & $17.5^{+0.5}_{-3.0}$ & [1500] & $1350^{+50}_{-60}$\\
Halo 2 & [0.00] & [0.00] & [0.13] & [-89.0] & $2.62^{+0.11}_{-0.81}$ & [45.89] & $680\tablenotemark{a}$\\
Halo 3 & [$-$0.21] & [$-$1.98] & [0.43] & [-23.7] & [0.16] & [41.60] & $50^{+70}_{-5}$\\
Halo 4 & [0.86] & [$-$2.87] & [0.01] & [24.3] & [0.07] & [19.13] & $100^{+50}_{-30}$\\ 
\cutinhead{Model 4} 
Halo 1 & $0.58^{+0.03}_{-1.96}$ & $7.64^{+1.67}_{-1.27}$ & $0.71^{+0.10}_{-0.01}$ & $109.7^{+7.7}_{-1.1}$ & $9.8^{+5.5}_{-1.1}$ & [1500] & $740^{+240}_{-70}$\\
Halo 2 & [0.00] & [0.00] & [0.13] & [-89.0] & $2.57^{+0.04}_{-0.67}$ & [45.89] & $660^{+20}_{-70}$\\
Halo 3 & [$-$0.21] & [$-$1.98] & [0.43] & [-23.7] & [0.16] & [41.60] & $80^{+30}_{-40}$\\
Halo 4 & [0.86] & [$-$2.87] & [0.01] & [24.3] & [0.07] & [10.13] & $90^{+40}_{-20}$\\
Halo 5 & $-0.96^{+3.74}_{-6.58}$ & $-18.76^{+2.81}_{-3.83}$ & $0.70^{+0.01}_{-0.15}$ & $143.0^{+5.7}_{-1.5}$ & $46.6^{+2.3}_{-6.3}$ & [1800] & $1800^{+40}_{-140}$\\
\enddata
\tablecomments{Values in brackets were held fixed during fitting. Halos 2, 3, and 4 are galaxy scale. They are labeled in cyan in Figure~\ref{zoom}.  Halo 5 takes into account the foreground structure, although it is projected to the same redshift as \spt, and thus the velocity dispersion is not indicative of its mass.  $\Delta$ RA and $\Delta$ DEC are measured in the image plane.  The ellipticity, $\epsilon$, is that of the mass distribution, while $\theta$ is the position angle of the potential, measured counter-clockwise from horizontal.}
\end{deluxetable*}

\section{Discussion and Conclusion} \label{conc}

Based on statistical tests, Model 1 is considered the best-fitting model.  We show the critical curves for this model for two different redshifts in Figure~\ref{fig:critic}.  The analysis that follows is based solely on Model 1.  It is also the model that is available through MAST.

\subsection{Strong Lensing Mass}

\begin{figure*}
\begin{center}
\subfigure{\includegraphics[scale=0.45]{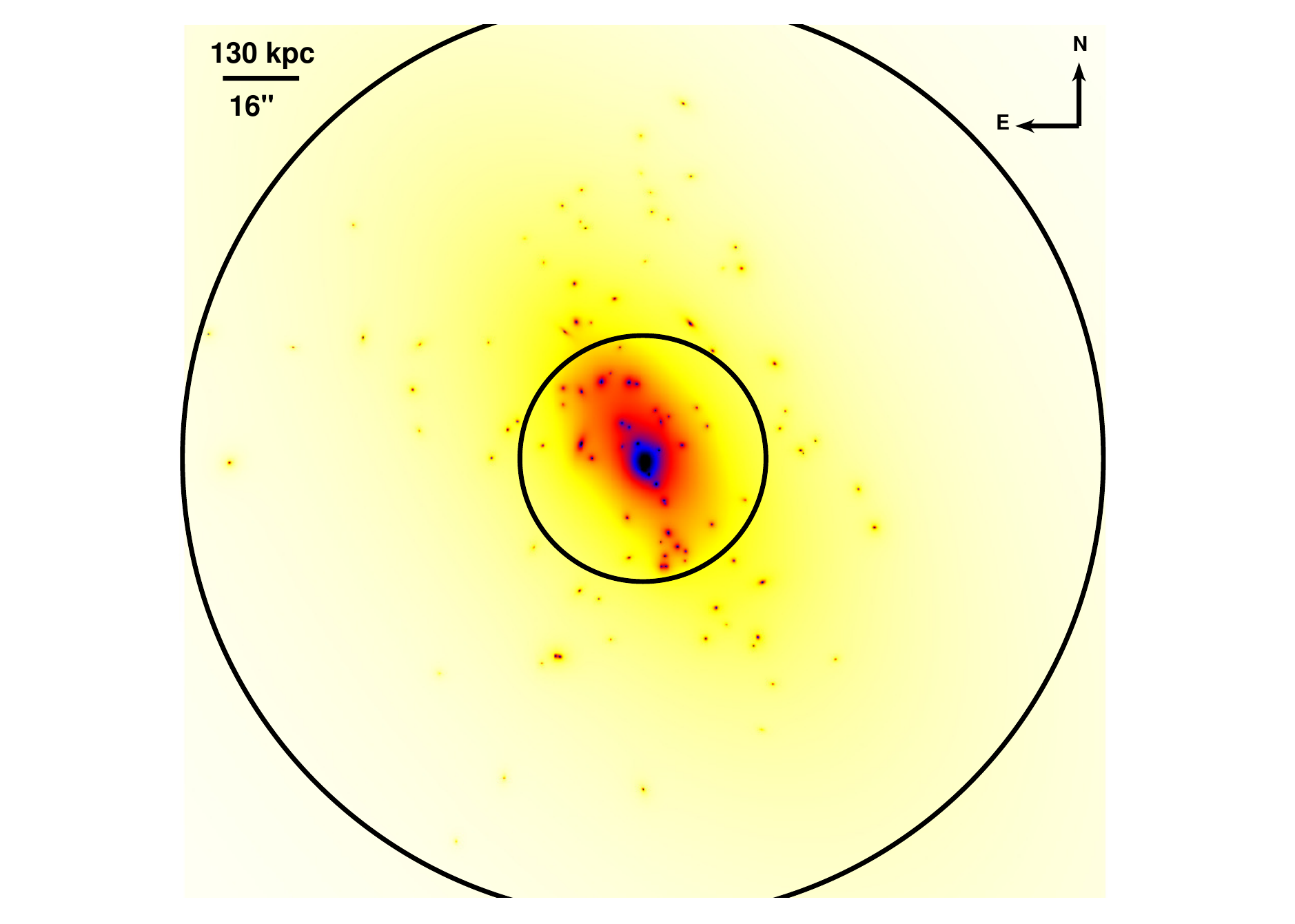}}
\subfigure{\includegraphics[scale=0.6]{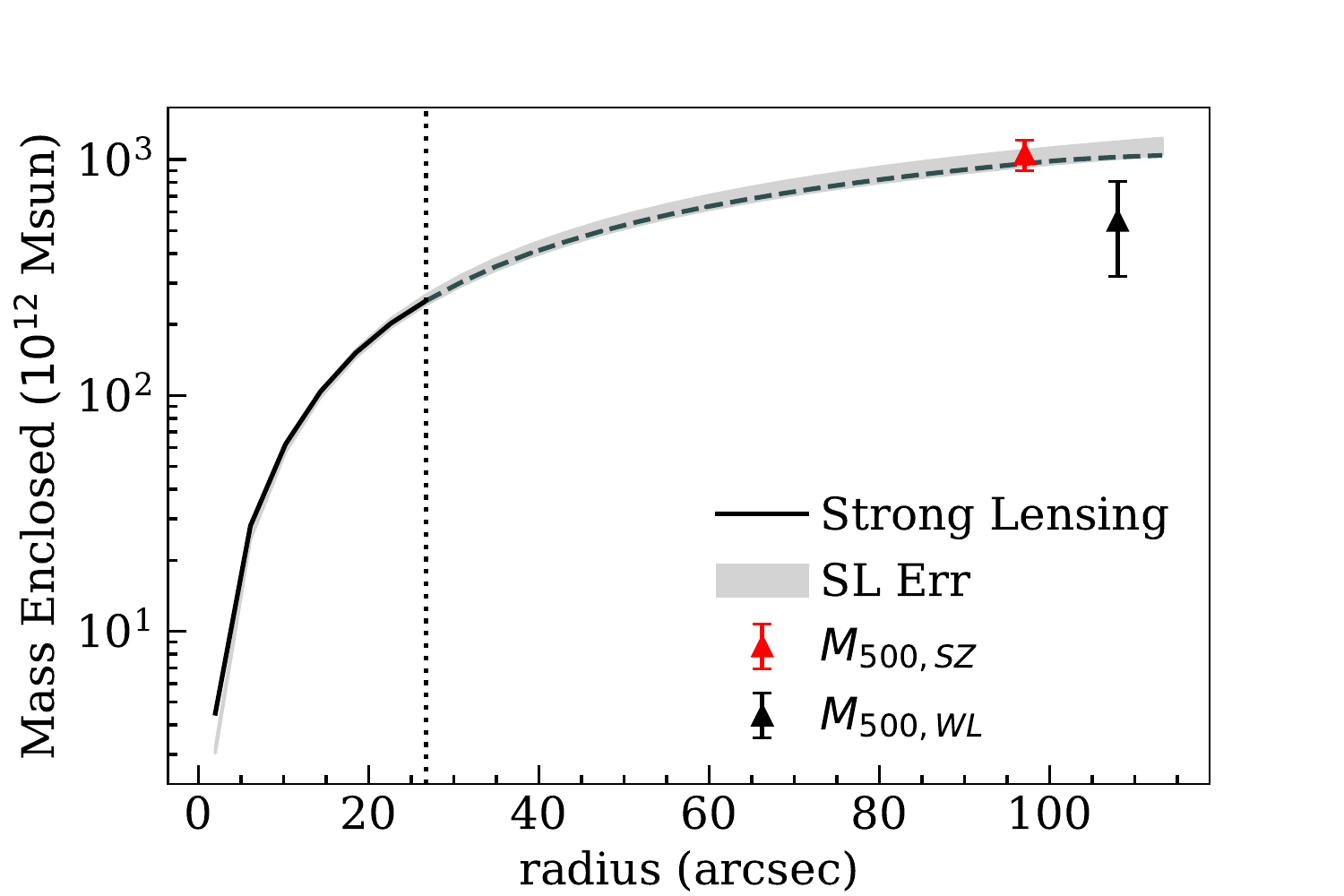}}
\caption{Left panel:  Mass map generated by lenstool.  Inner annulus is at $r=26.72''$, which is the limit of our strong lensing constraints.  Outer annulus is at $r=100''$, which corresponds to $r_{500}$.  Right panel:  Projected mass enclosed within radius $r$.  The solid black line shows the results from our model.  The gray shaded region shows the error.  The dashed grey line shows the region where there are no strong lensing constraints and the mass enclosed is extrapolated.  Red and black triangles show $M_{500}$ calculated from SZ and weak lensing, respectively.  The weak lensing mass has not been corrected for bias. \edit1{As noted in the text, the weak lensing mass reported here is a spherical overdensity mass, which tends to be lower than the enclosed mass given the projection, and thus the measurement is consistent with our extrapolated strong lensing mass. }}
\label{massfig}
\end{center}
\end{figure*}

We calculate the projected mass of the cluster using the mass map generated by \texttt{Lenstool} (Figure~\ref{massfig}, left).  To calculate the $1\sigma$ error bars, we generate 100 maps from parameter sets sampled from the MCMC analysis and calculate the standard deviation of the distribution of calculated masses.  Strong lensing mass calculations are most accurate in the region where there are constraints.  Our convention is as follows:  we use $R$ for the 2D projected radius and $r$ for the 3D spherical radius.  For \spt, there are constraints out to $R\sim25''$.  We find the total projected mass density within $R=26.7''$ to be $M=2.51^{+0.15}_{-0.09}\times 10^{14}$~M$_{\sun}$.  We also extrapolate a mass measurement to $R_{500}$ (the dashed black line in Figure~\ref{massfig}, right) so that we may compare to other studies of this cluster.  
In particular, we compare our constraints to the weak lensing analysis conducted by \citet{schrabback}, which is based on the mosaic ACS
observations of the cluster.
When centering their weak lensing measurements onto the {\it Chandra} X-ray centroid and correcting for the corresponding miscentring and
mass modelling bias, these authors constrain the cluster mass to 
$M_{500,WL}=5.5^{+2.6}_{-2.3}\times 10^{14}\mathrm{M}_\odot$.
Assuming the \citet{diemer15} concentration--mass relation their best-fitting mass  corresponds to a spherical overdensity radius of $r_{500,WL}=108^{\prime\prime}$.
This WL mass constraint agrees within \mbox{$2\sigma$}  with the SZ constraint  
\mbox{$M_{500,SZ}=10.53\pm 1.55\times 10^{14}\mathrm{M}_\odot$},
which \citet{bleem} obtain when assuming a mass-observable scaling relation for which
the SPT cluster counts fit a $\Lambda$CDM cosmology best
\citep{reichardt13}.
We compare these estimates of the spherical overdensity mass, plotted at the
WL-estimated $r_{500,WL}$, to the enclosed mass from our
extrapolated SL model in the right panel of Figure~\ref{massfig}.
Noting that the enclosed mass at $R_{500}$ is generally higher
than a spherical overdensity mass  at $r_{500}$ given the projection, we
conclude that the different mass measurements are broadly consistent.
Again, we emphasize that we are unable to constrain the mass slope with strong lensing this far outside the region of the strong lensing constraints. The statistical errors grossly underestimate the true uncertainties at these projected radii, and thus these estimates should be used with caution.

At $M\sim10^{15}$~M$_{\sun}$, \spt is one of the most massive high-redshift clusters known.  The only other cluster in the RELICS sample with $z>0.7$ is ACT-CLJ0102$-$49151 (```El Gordo'').  It is at $z=0.870$ and has $M_{200, SZ} = 2.16\pm0.32 \times 10^{15}~h^{-1}_{70}$~M$_{\sun}$~\citep{elgordo}.  A strong lensing analysis by \citet{zitrin} found a lower limit of $M\sim1.7\times10^{15}$~M$_{\sun}$, in good agreement with the SZ mass.  The strong lensing analysis by \citet{cerny} finds that $M(<500~\textrm{kpc})=11.0\pm0.7\times10^{14}$~M$_{\sun}$, also in good agreement.  Other strong lensing clusters with complete models in this high-redshift regime include RCS 0224-0002 ($z=0.773$, \citealp{gladders2002, smit}) with $M_{200, SL} = 1.9\pm0.1\times 10 ^{14}$~M$_{\sun}$~\citep{rzepecki}, and RCS2 J232727.6-020437 ($z=0.7$, \citealp{gilbank2011, hoag, menanteau13}) with $M_{200}\sim 2-3\times 10^{15}h_{70}^{-1}\mathrm{M}_\odot$~\citep{sharon,schrabback18b}.  High-redshift clusters that show evidence of strong lensing but do not have complete models include RCS 231953$+$0038.0 ($z=0.897$, \citealp{gladders2002}) and IDCS J1426.5+3508 ($z=1.75$, \citealp{gonzalez}).  RCS 231953$+$0038.0 is part of a supercluster, along with two other cluster components~\citep{gilbank2008}.  It has an X-ray mass of $M_{200, X}= 6.4^{+1.0}_{-0.9}\times 10 ^{14}$~M$_{\sun}$~\citep{hicks, gilbank2008} and a weak-lensing mass of $M_{200, WL} = 5.8^{+2.3}_{-1.6}\times 10^{14}$~M$_{\sun}$~\citep{jee}.  The cluster IDCS J1426.5+3508 is the most massive cluster known at $z>1.4$.  \citet{gonzalez} use the presence of a giant strong lensing arc to calculate the cluster mass enclosed within the arc.  Extrapolating, they find $M_{200, SL} > 2.8^{+1.0}_{-0.4}\times 10 ^{14}$~M$_{\sun}$.  Comparing \spt to the other known strong-lensing clusters at high redshift, we conclude that it is not a mass outlier in the group of known strong-lensing clusters.

The high mass of \spt is likely a contributing factor to its success as a lensing cluster, as it has the second highest number of high-redshift ($z>5.5$) galaxy candidates in the RELICS sample.  El Gordo also has a significant number of high-redshift candidates, coming in fourth in the RELICS sample~\citep{salmon}.  While a systematic search for high-redshift galaxy candidates has not been undertaken for the other clusters mentioned in this section, it is likely that the combination of the their high mass and high-redshift combine to make them good candidates for searching for high redshift galaxy candidates in their fields.

\subsection{The Presence of a $z\sim10$ Arc} \label{hizarc}

SPT0615-JD is a candidate $z\sim10$ ($z_{phot}=9.9\pm0.6$) galaxy gravitationally lensed into an arc spanning $2\farcs5$ in the field of \spt.  It was found as part of a systematic search for high-redshift galaxies in the RELICS fields~\citep{arc}.  It is not visible in bands blueward of F140W.

\begin{figure*}
\begin{center}
\subfigure{\includegraphics[scale=0.41]{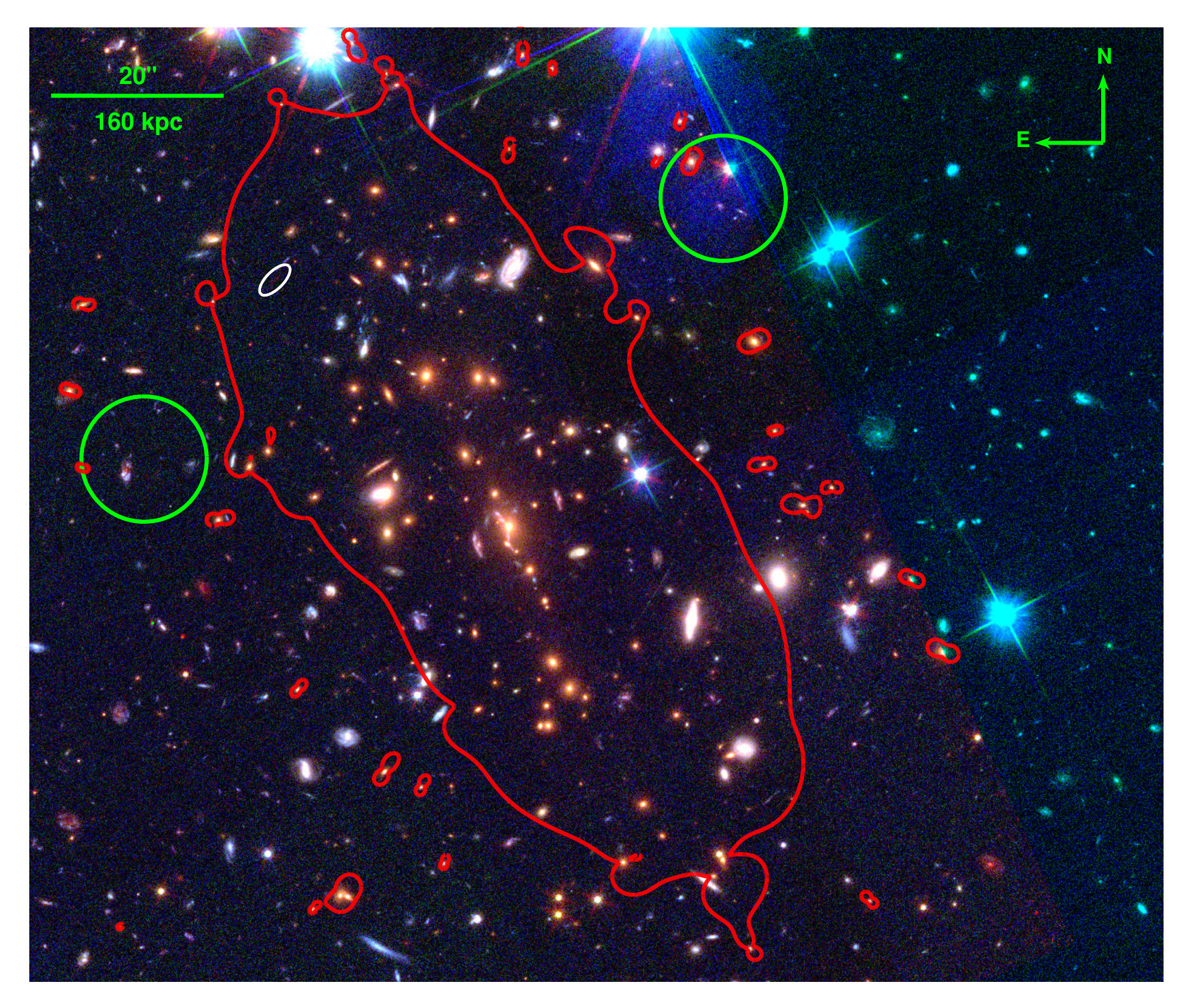}}
\subfigure{\includegraphics[scale=0.43]{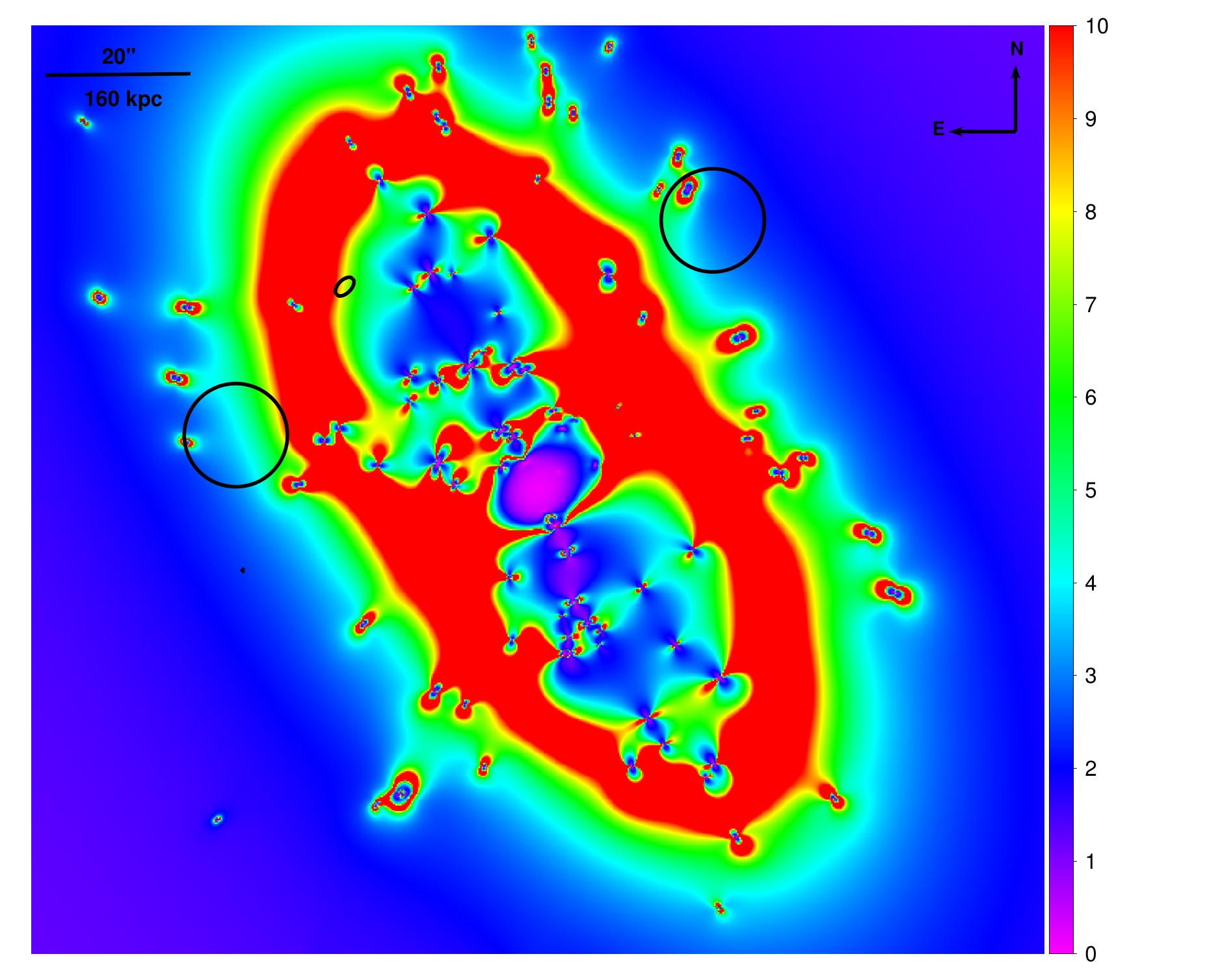}}
\caption{Left panel:  Same as Figure~\ref{fig:critic}, but with the high-redshift candidate galaxy marked in white and predicted locations for the multiple images marked in green.  One of the predicted images is next to a bright star, so will be difficult to see.  Right panel:  Magnification map for a source at $z=9.93$.  The high-redshift candidate galaxy and the predicted locations for multiple images are in black.  Regions with $\mu\ge1$ are magnified.  SPT0615-JD is magnified by $\sim8\times$. Both of the predicted locations have magnifications ranging from 3-6$\times$.  }
\label{hizfig}
\end{center}
\end{figure*}

The left panel of Figure~\ref{hizfig} shows the location of this galaxy, along with the predicted locations of counterimages.  The right panel shows the magnification map produced by our lens model for $z=9.9$.  The counter-image in the upper-right hand corner is predicted to be $\sim1$~magnitude fainter than the original arc, placing it below the detection level of HST.  Its location next to a large star also makes it difficult to search for.

Using our best-fit model, the counter-image in the east is predicted to be 0.04 magnitudes fainter than SPT0615-JD, which should be visible at the depth of our images; however a search in that region has not yielded a counter-image.  The arc is aligned with the direction of the shear.  We note that all the models predict counterimages in the same location and with approximately the same \edit1{magnification}, with the exception of Model 3, which only predicts one counterimage to the northwest.  A \texttt{GLAFIC} model (\citealp[]{oguri}, Kikuchihara et al., in preparation) and Light Traces Mass~\citep{ltm} model both predict counterimages in the same location (see \citet{arc} for more details).  The right panel of Figure~\ref{hizfig} shows that SPT0615-JD is magnified by $\sim8\times$ the intrinsic brightness, while the predicted counter-image would be magnified by $3-6\times$ the intrinsic brightness of the galaxy.       

\subsection{Conclusion}

We present a strong lens model for the cluster SPT-CLJ0615$-$5746 (also known as PLCKG266.6$-$27.3) based on the presence of three multiply imaged background galaxies.  Two of these multiply imaged families have confirmed spectroscopic redshifts from our observations with Magellan.  The best model using the statistical results from the BIC and \aic is Model~1, which optimizes one cluster-sized dark matter halo and three smaller galaxy-sized haloes, in addition to cluster-member galaxies whose mass is determined from their light through scaling relations.  This model only includes the secure observations of system~3, as well as the secure images from families 1 and 2.  There are additional predicted images of system 3; however these need spectroscopic confirmation before including them in the model.   

The lens model is complicated by the presence of a foreground structure, estimated to be at a photometric redshift $z\sim0.4$.  This is not surprising, given the prevalence of line-of-sight structure~\edit1{\citep{bayliss}}.  We made versions of the lens model including this foreground structure, but the statistical analysis did not favor either version.  Our analysis was not a full multiplane analysis, however, which is currently not fully supported by \texttt{Lenstool}.  Such analysis would also benefit from spectroscopic confirmation of both the foreground candidates and multiply-imaged background galaxies.  

\spt is a massive high-redshift cluster, with a strong-lensing mass of $M_{500}=10.62\pm0.77\times10^{14}$~M$_{\sun}$.  Our strong lensing mass is comparable to the SZ determined mass.  It is similar in mass to other strong lensing clusters in the $z>0.8$ regime, and has been shown to have magnified a high number of high-redshift background galaxies into our detection limit~\citep{salmon}.  The field also contains a high-redshift galaxy candidate with a photometric redshift $z=9.93$~\citep{arc}.  

\spt is included in the RELICS program, and as such the data for this lens model are available through MAST.  This data includes reduced images, catalogs, and lens models.

\facilities{HST, Magellan}

\acknowledgements
Support for program GO-14096 was provided by NASA through a grant from the Space Telescope Science Institute, which is operated by the Association of Universities for Research in Astronomy, Inc., under NASA contract NAS5-26555. This paper is based on observations made with the NASA/ESA Hubble Space Telescope, obtained at the Space Telescope Science Institute, which is operated by the Association of Universities for Research in Astronomy, Inc., under NASA contract NAS 5-26555. These observations are associated with program GO-14096. Archival data are associated with programs GO-12757 and GO-12477. This paper includes data gathered with the 6.5 meter Magellan Telescopes located at Las Campanas Observatory, Chile.

\bibliographystyle{apj}

\bibliography{spt0615}

\end{document}